\documentclass[prl,twocolumn,nofootinbib,longbibliography]{revtex4-1}
\usepackage{graphicx}
\usepackage{color}
\usepackage{epsfig}
\usepackage[caption=false]{subfig}

\begin{document}

\title{Block Analysis for the Calculation of Dynamic and Static Length Scales in Glass-Forming Liquids}
\author{Saurish Chakrabarty$^{1}$}\thanks{These authors contributed equally.}
\author{Indrajit Tah$^{2}$}\thanks{These authors contributed equally.}
\author{Smarajit Karmakar$^{2}$}
\email{smarajit@tifrh.res.in}
\author{Chandan Dasgupta$^{3,4}$}
\email{cdgupta@physics.iisc.ernet.in}
\affiliation{
  $^1$ International Centre for Theoretical Sciences,
  Tata Institute of Fundamental Research, 
  Shivakote, Hesaraghatta, Hubli, Bangalore, 560089, India,\\
  $^2$ Centre for Interdisciplinary Sciences,
  Tata Institute of Fundamental Research, 
  21 Brundavan Colony, Narisingi, Hyderabad, 500075, India,
  $^3$ Centre for Condensed Matter Theory, Department of Physics,
  Indian Institute of Science, Bangalore, 560012, India, 
  $^4$ Jawaharlal Nehru Centre for Advanced 
  Scientific Research, Bangalore 560064, India.}

\begin{abstract}
We present {\it block analysis}, an efficient method to perform finite-size scaling for 
obtaining the length scale of dynamic heterogeneity and the point-to-set length scale 
for generic glass-forming liquids.  This method involves considering blocks of varying 
sizes embedded in a system of a fixed (large) size. The length scale associated with 
dynamic heterogeneity is obtained from a finite-size scaling analysis of the dependence 
of the four-point dynamic susceptibility on the block size. The block size dependence 
of the variance of the $\alpha$-relaxation time yields the static  point-to-set length 
scale. The values of the obtained length scales agree quantitatively with those obtained 
from other conventional methods.  This method provides an efficient experimental tool 
for studying the growth of length scales in systems such as colloidal glasses for which 
performing finite-size scaling by carrying out experiments for varying system sizes may 
not be feasible.
\end{abstract}

\maketitle

The role of growing length scales in the rapid growth of the structural relaxation time 
of glass-forming liquids near the glass transition~\cite{book1,book2} has received a lot 
of attention in recent years~\cite{rev1,rev2}. Several length scales, both static and 
dynamic, have been proposed~\cite{rev1,rev2} and their behavior near the glass transition 
and relevance to the growth of the structural relaxation time have been studied in a 
large number of theoretical~\cite{11BB,ARPC58,PhysRevLett.97.195701,ktw4,JCP121-7347}, 
numerical~\cite{PNASUSA2009,PhysRevLett.105.015701,IT,PhysRevLett.105.217801,Nature2008,
Hocky2012,kobVargasBerthier,szamel3,PhysicaA,PhysRevLett.111.165701,CTPRE12} and 
experimental~\cite{annurev.physchem.51.1.99,colloid,kegel,05Berthier,dhlengthexp2,
fifthorder,PhysRevLett.116.098302,rajesh} investigations. However, there is still a lot 
of controversy about the behavior of these length scales as the glass transition is 
approached. Therefore, it is important to develop methods that can be used in simulations 
and experiments to accurately measure these length scales.

The existence of a growing dynamic length scale $\xi_D$ that describes spatial correlations 
of the inhomogeneous local dynamics of glass-forming liquids (known as dynamic 
heterogeneity~\cite{annurev.physchem.51.1.99}) is now well-established. This
length scale has been calculated from finite-size scaling~\cite{PNASUSA2009} (FSS) of a 
four-point susceptibility $\chi_4$~\cite{cdg1} and its associated Binder cumulant~\cite{fss} 
and the wavenumber-dependence of the corresponding structure factor ~\cite{PhysRevLett.105.015701,PhysRevLett.105.217801}. A similar length scale has also 
been obtained~\cite{kobVargasBerthier} from the dependence of the local dynamics on the 
distance from an amorphous wall in which particles are fixed at their positions in an 
equilibrium configuration. However, the relation of this length scale with the length 
scale $\xi_D$ of dynamic heterogeneity is controversial~\cite{szamel3}. Inhomogeneous 
mode-coupling theory~\cite{PhysRevLett.97.195701} provides a theoretical description 
of the growth of $\xi_D$ and $\chi_4$ as the glass transition is approached, but the 
quantitative predictions of this theory are somewhat different from the results obtained 
from numerical studies~\cite{PNASUSA2009,PhysRevLett.105.015701,PhysRevLett.105.217801}.

Another length scale that has received a lot of attention is the static ``mosaic scale'' 
($\xi_s$) of the Random First-Order Transition (RFOT) theory~\cite{ARPC58,ktw4} of the 
glass transition. This length scale can be obtained~\cite{JCP121-7347} from a 
``point-to-set'' (PTS) construction in which particles outside a spherical cavity 
are fixed at their positions in an equilibrium configuration, the remaining particles 
inside the cavity are allowed to equilibrate, and the average overlap of the positions 
of these particles with their positions in the original equilibrium configuration is 
studied as a function of the radius of the cavity. The PTS method has been used in 
several studies~\cite{Nature2008,Hocky2012,CTPRE12} to obtain the dependence of 
$\xi_s$ on the temperature and the density. This length scale has also been calculated 
from FSS of the $\alpha$-relaxation time~\cite{PNASUSA2009} and the minimum eigenvalue 
of the Hessian matrix~\cite{PhysicaA,PhysRevLett.111.165701} that describes vibrations 
near a local minimum of the potential energy. In the temperature and density range 
accessible in simulations, $\xi_s$ is found to be smaller than $\xi_D$ and the growth 
of $\xi_s$ with decreasing temperature or increasing density does not  follow that of 
$\xi_D$, suggesting that these two length scales are distinct from each other.

Experimental studies of length scales in glass-forming liquids have been limited 
because quantities such as $\chi_4$ that are required for calculating these length 
scales are not readily accessible in experiments. A calculation of $\chi_4$ requires 
information about the trajectories of individual particles, which can be obtained in 
experiments on colloidal systems~\cite{colloid,kegel}, but not in experiments on 
molecular liquids. Three-point and five-point susceptibilities that are closely 
related to $\chi_4$ have been measured in experiments 
~\cite{05Berthier,dhlengthexp2,fifthorder} on molecular liquids. These experiments 
and experiments on colloidal systems~\cite{colloid,kegel} provide clear evidence for 
the growth of  spatial correlations as the glass transition is approached. However, 
it is difficult to extract values of relevant length scales from these measurements 
because the exact relation between these susceptibilities and the length scales is 
not known. A calculation of $\xi_s$ using the PTS method requires detailed information 
about the equilibrium properties of particles confined in small cavities of varying 
sizes. Such information is difficult to obtain from experiments, although a recent 
experiment~\cite{PhysRevLett.116.098302} suggests that this may be possible in the near 
future. The only experiment in which values of both dynamic and static length scales 
have been obtained is Ref.~\cite{rajesh} in which the method of 
Ref.\cite{kobVargasBerthier} was implemented for a two-dimensional colloidal system. 
However, as mentioned earlier, the physical interpretation of the length scales 
obtained from this procedure is controversial.

As discussed above, FSS has played an important role in the calculation of both 
$\xi_D$~\cite{PNASUSA2009} and $\xi_s$~\cite{PNASUSA2009,PhysicaA,PhysRevLett.111.165701}. 
The conventional FSS method involves studies of the equilibrium behavior of systems with 
periodic boundary conditions and different sizes that are comparable to the (often rather 
small) values of the relevant length scales. This method can not be implemented in 
experiments because experimental studies of such systems are very difficult. It is, 
therefore, important to develop alternative FSS methods that can be implemented in 
experiments. Also, the conventional FSS method suffers from a few problems such as the 
necessity of carrying out long simulations and extensive averaging for obtaining reliable 
results for small systems and artifacts\cite{PhysRevLett.105.217801} arising from the 
suppression of density and composition fluctuations in small samples with periodic boundary 
conditions. FSS methods in which these problems are not present would be highly desirable. 

In this Letter, we present a method of performing FSS in which the problems of 
conventional FSS analysis are avoided, leading to excellent scaling behavior.  In this 
method, which we call ``block analysis'', we perform equilibrium molecular dynamics (MD) 
simulations for a single large system.  We then consider blocks of varying sizes embedded 
in the large system~\cite{blocking} and measure various quantities of interest, such 
as $\chi_4$ and the $\alpha$-relaxation time $\tau_\alpha$, for individual blocks.  
The length scale $\xi_D$ associated with dynamic heterogeneity is obtained from a FSS 
analysis of the dependence of  $\chi_4$ and the associated Binder cumulant on the 
block size. The block-size dependence of the variance of the $\alpha$-relaxation time 
of individual blocks yields the static PTS length scale $\xi_s$. We show that the 
values of the obtained length scales agree quantitatively with those obtained from 
other methods for three different glass-forming liquids. Since this method involves 
observation of the trajectories of particles in a single large system, it can be readily 
implemented in experiments on colloidal systems.
 
We study three model glass-formers in three spatial dimensions. The first is the 
well-known Kob-Andersen binary mixture interacting via Lennard-Jones potentials 
~\cite{KA}. We call this system the 3dKA model. Second, we study 
a $50:50$ binary mixture interacting via purely repulsive interactions falling of 
as $1/r^{10}$ ~\cite{R10}. We call this the 3dR10 model. Lastly, we study 
a variant of 3dKA system with only repulsive power law interactions
~\cite{IPL}. We refer to this as the 3dIPL model. Further details of the 
models and simulations can be found in the supplementary information (SI).

\begin{figure*}[htpb]
  \begin{center}
    \includegraphics[scale=0.4]{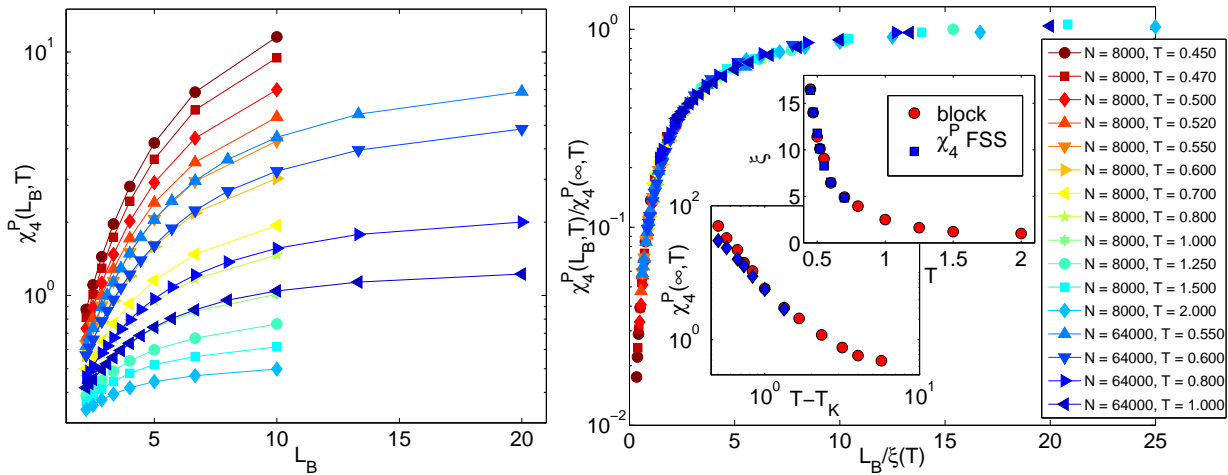}
     \vskip -0.3cm
    \caption{Block size dependence of $\chi_4^P$
      for the 3dKA
      model. In the left 
      panels, $\chi_4^P$ is plotted against the block size
      and in the right panel a collapse of the data is obtained by rescaling the
      $x$-axis using a suitable length-scale,
      $\xi(T)$ and the $y$-axis using the
      saturation value of $\chi_4^P$ for infinite block size.
      In the insets, the temperature dependences of $\xi(T)$ and
      $\chi_4^P(\infty,T)$ are shown and compared with the
      corresponding quantities obtained using conventional FSS.
      For 3dKA model we have used $T_K \simeq 0.30$.}
    \label{chi4BlockAnalysis3dKAR10}
  \end{center}
  \vskip -0.7cm
\end{figure*}
MD simulations are carried out for a single, moderately large 
system size, $N=\rho L_0^3$, where $\rho$ is the number density and 
$L_0$ is the length of the system. We then construct blocks of size $L_B=L_0/n$, 
where $n\in\{2,3,4,~.~.~.\}$ and calculate various dynamic quantities
using the particles which are present inside one such box at a chosen
time origin. 

\vskip +0.5cm
\noindent{\bf The dynamic susceptibility $\chi_4$:}
The self overlap correlation for a particular block size is defined as,
\begin{eqnarray}
  Q(L_B,t)=
  \frac{1}{N_B}
  \sum_{i=1}^{N_B}
  \frac{1}{n_i}
  \sum_{j=1}^{n_i}
  w(|\vec{r}_j(t)-\vec{r}_j(0)|),
  \label{qlt}
\end{eqnarray}
where $N_B$ is the number of blocks with size $L_B$, $n_i$ is the number
of particles in the $i$-th block at time $t=0$,
and the window function $w(x)=\Theta(a-x)$ where $\Theta$ is
the Heaviside step function and the value of the parameter $a$ is chosen 
to remove the decorrelation arising from vibrations of particles inside 
the cages formed by their neighbours. We take $a$ to be $0.3\sigma_{AA}$ 
for the 3dKA model where $\sigma_{AA}$ is the Lennard-Jones
length parameter for the larger particles. The dynamical susceptibility 
associated with blocks of size $L_B$ is defined as follows.
\begin{eqnarray}
  \chi_4(L_B,t)=
  \frac{NL_B^3}{L_0^3}
  \langle
  [Q(L_B,t)-
  \langle
  Q(L_B,t)
  \rangle]^2
  \rangle
\end{eqnarray}

%
We consider the dependence 
of $\chi_4^P(L_B,T)$, the peak value of $\chi_4(L_B,t)$ at temperature $T$, 
on the block size $L_B$ for a fixed value of $N=\rho L_0^3$. This dependence 
is shown in Fig. ~\ref{chi4BlockAnalysis3dKAR10}. The left panel of the figure shows 
the data for $\chi_4^P(L_B,T)$ as a function of the block length $L_B$ for
different temperatures. The peak value of the dynamical susceptibility at a 
given temperature grows with $L_B$ and saturates at a temperature-dependent value
$\chi_4^P(\infty,T)$. The dependence of $\chi_4^P(L_B,T)$ on $L_B$ is expected
to exhibit the following FSS form:
\begin{equation}
\chi_4^P(L_B,T) = \chi_4^P(\infty,T) f(L_B/\xi(T)),
\label{fsschi4}
\end{equation}
with $\xi(T) = \xi_D(T)$, the dynamic length scale. The data for all temperatures 
can be collapsed to a master curve using the two parameters, $\chi_4^P(\infty,T)$ 
and $\xi(T)$, for each temperature, as shown in the right panel of 
Fig. ~\ref{chi4BlockAnalysis3dKAR10}. The quality of the data collapse is very 
good and the length scale obtained this way is in complete agreement with that 
obtained using conventional FSS, as shown in the inset of the same figure. The 
legend ``$\chi_4^P$ FSS'' refers to conventional FSS and the data are taken from
Ref.\cite{PNASUSA2009}. We have also shown the comparison of $\chi_4^P(\infty,T)$
with the conventional FSS values. One can see that at low temperatures,
$\chi_4^P(\infty,T)$ obtained from the block analysis is systematically larger than
the conventional FSS result. This is due to the fact that particles can move
in and out of individual blocks, so that the constraint of the total number of 
particles of each type being constant in simulations with periodic boundary 
conditions is not present for the blocks.
This enhances fluctuations, leading to an increase in the peak height of $\chi_4(t)$ for the blocks.

To ascertain whether the above analysis generically gives correct results for
any model system, we have performed similar analysis for the
3dIPL and 3dR10 model systems. For both these models, the scaling collapses observed
are quite good and the extracted dynamic length scales are also in good 
agreement with those obtained from conventional methods (see the SI for details).

\begin{figure*}
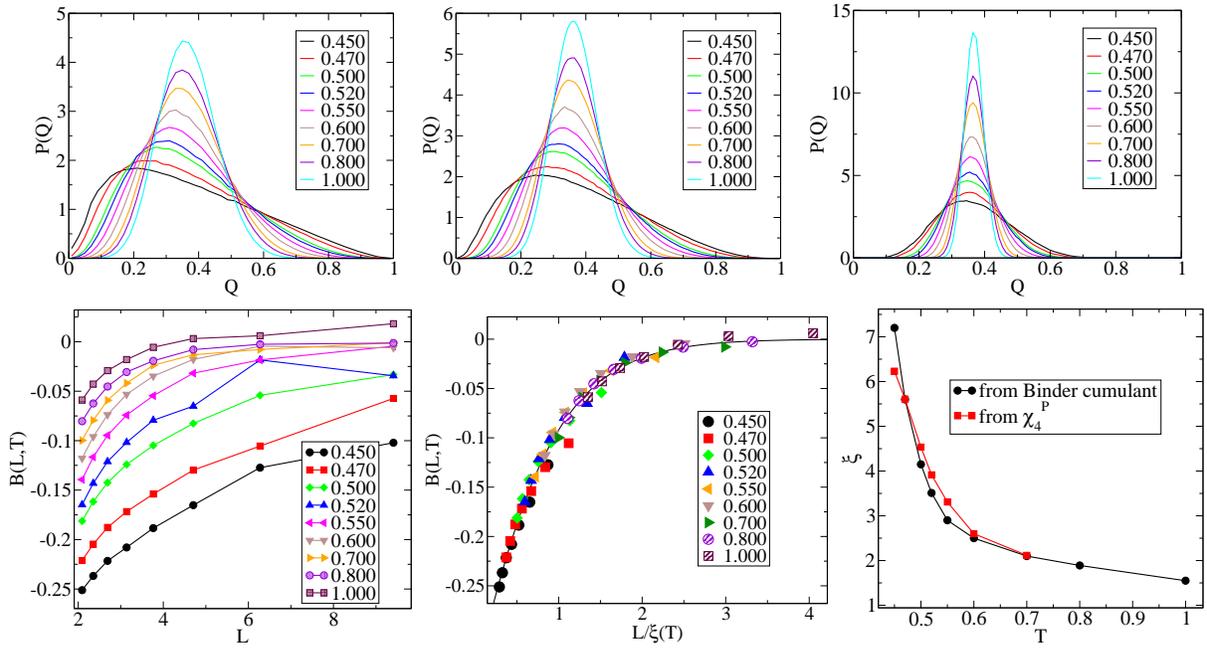

  \begin{center}
    \includegraphics[width=0.270\textwidth]{histQl3_764.eps}\label{h3.764}
    \hskip +0.2cm
    \includegraphics[width=0.270\textwidth]{histQl4_705.eps}\label{h4.705}
    \hskip +0.2cm
    \includegraphics[width=0.280\textwidth]{histQl9_410.eps}\label{h9.410}
\\
\vskip +0.1cm
\hskip -0.1cm
\includegraphics[width=0.3\textwidth]{bc.eps}
\hskip +0.05cm
\includegraphics[width=0.3\textwidth]{bcCollapse2.eps}
\hskip +0.05cm
\includegraphics[width=0.27\textwidth]{bcLength2.eps}
    \caption{(For the 3dKA model) {\em Top panel:} Histograms of $Q(L_B,\tau_\alpha)$ for blocks of size
      $L_B=$ 3.764, 
      6.274 and 9.410, going from left to
      right.
      {\em Bottom left panel:} Binder cumulant versus block size at different temperatures.
      {\em Bottom middle panel:} Binder cumulant versus
      block size scaled using the lengths shown in the {\em bottom
        right panel} where the length scale used for the collapse of
      $\chi_4^P$ data has also been included for comparison.
    }
    \label{histQ}
  \end{center}
  \vskip -0.5cm
\end{figure*}

\vskip +0.5cm
\noindent{\bf Distribution of $Q(\tau_\alpha)$ and the Binder Cummulant:}
The FSS of $\chi_4^P(N,T)$ requires two unknown parameters, $\chi_4^P(\infty,T)$ 
and $\xi(T)$. A better way of extracting the length scale is the FSS of the Binder 
cummulant obtained from  the distribution of $Q(\tau_\alpha)$ where $\tau_{\alpha}$
is the long time $\alpha$-relaxation time defined as 
$\langle Q(L_0,t=\tau_{\alpha})\rangle = 1/e$. 
At a fixed temperature, the distribution of $Q(\tau_\alpha)$ becomes flatter and more
skewed as $L_B$ is decreased, becoming nearly bimodal for small $L_B$.
A similar effect is seen if the temperature is decreased for fixed $L_B$. This is 
because lowering the temperature is another way of lowering the block size measured 
in units of the dynamic correlation length. This is clearly seen in the top
panels of Fig.~\ref{histQ}.
An earlier study~\cite{PNASUSA2009} reported that the distribution of $Q(\tau_\alpha)$ 
becomes bimodal for small systems with periodic boundary conditions. Such bimodality is 
not observed in our analysis using the block construction, presumably because of 
enhanced fluctuations of the density and the composition in the blocks for which the 
number of particles of each type is not conserved.

The skewness of the distribution of $Q(\tau_\alpha)$ is quantitatively
captured by the Binder cumulant which measures the deviation of the distribution 
from the Gaussian form. It is define as 
\begin{eqnarray}
  B(L_B,T)=\frac{\langle [Q(L_B,\tau_\alpha)-\langle Q(L_B,\tau_\alpha)\rangle]^4\rangle}
  {3\langle [Q(L_B,\tau_\alpha)-\langle Q(L_B,\tau_\alpha)\rangle]^2\rangle^2}
  -1.
\end{eqnarray}
This quantity approaches zero at high temperatures and for large block sizes where 
the correlation length in the system is much smaller than the block 
size. The Binder cummulant is an ideal
quantity to measure in a scaling analysis because it is known to be a scaling function
of only the underlying correlation length:  
\begin{equation}
B(L_B,T) = \mathcal{F}\left(\frac{L_B}{\xi_D}\right). 
\end{equation}
The estimation of the length scale from FSS of $B(L_B,T)$
is more reliable as it involves only one parameter for each 
temperature. The bottom left panel of Fig. \ref{histQ} shows the Binder cumulant 
calculated from the distribution of $Q(L_B,\tau_\alpha)$,
plotted versus $L_B$ for the 3dKA model. The middle panel shows the corresponding
data collapse obtained using the dynamic length scale shown in the bottom right
panel of the same figure. We also show a comparison of the length scales
obtained using FSS of $\chi_4^P$ and the Binder cummulant. The length scales 
obtained in these two calculations are in reasonably good agreement with each other.



\vskip +0.5cm
\noindent{\bf The statistics of $\tau_\alpha$ - Calculation of the
  static length-scale:}
  \begin{figure}[htpb]
  \vskip -0.4cm
\begin{center}
\hskip -1.2cm
\includegraphics[width=0.6\textwidth]{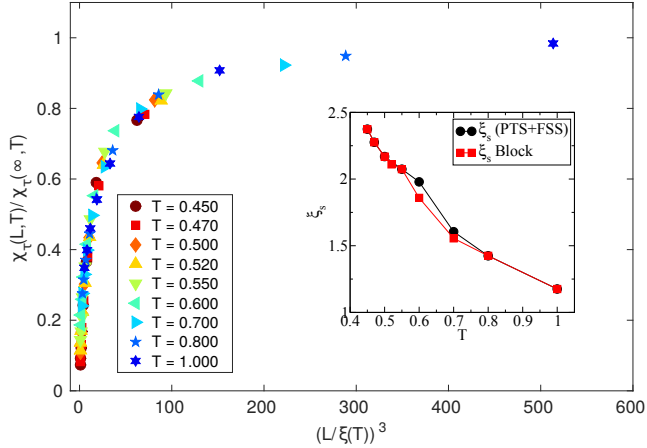}\hskip -1.2cm
\vskip -0.3cm
\caption{
(For the 3dKA model) Data collapse of block-size dependence of $\chi_{\tau}$ (see Eq.(\ref{chitau})) for different 
temperatures using the static length scale. Inset shows the comparison of the static length
scales obtained from this method and from Ref.\cite{PhysRevLett.111.165701}.}

\label{tauChi}
\end{center}
\vskip -0.4cm
\end{figure}
Earlier studies~\cite{PNASUSA2009,PhysRevLett.111.165701} have shown that the $\alpha$-relaxation time of small systems with periodic boundary conditions increases as the system size is decreased and its system-size dependence is described by the static length scale $\xi_s$. We find that
the $\tau_\alpha$ for individual blocks, obtained from the self overlap
correlation function $Q(t)$, does not show appreciable block-size
dependence - the movement of particles in and out of blocks makes the system-size dependence of $\tau_\alpha$ for blocks much weaker than that for systems with periodic boundary conditions. We then look at the dependence of the statistics of 
$\tau_\alpha$ on the block size by calculating the 
distribution of $\tau_\alpha$ as a function of block size.
For each block, we first calculate $\tau_\alpha^{(i)}(L_B)$ by measuring the 
time at which the corresponding $Q^{(i)}(L_B,t)$ for a fixed time origin
attains a value of $1/e$ (the superscript ${(i)}$ signifies that this
is a quantity for a single block $i$ before any averaging is done). We then calculate the mean and the variance of this quantity and
finally define
$\chi_\tau(L_B,T)$ as,
\begin{equation}
\chi_\tau(L_B,T) =
L_B^3\left\langle
\frac{\frac{1}{n_B}\sum_{i=1}^{n_B}[\Delta\tau_{\alpha}^{(i)}(L_B)]^2}{\overline{ \tau_{\alpha}^{(i)}(L_B)}}\right\rangle,
\label{chitau}
\end{equation}
where $\overline{\tau_{\alpha}^{(i)}(L_B)} = 
\frac{1}{n_B}\sum_{i=1}^{n_B} \tau_{\alpha}^{(i)}(L_B)$, $\Delta \tau_\alpha^{(i)}(L_B) = \tau_\alpha^{(i)}(L_B)-\overline{\tau_{\alpha}^{(i)}(L_B)}$,
and the outermost angular brackets stand for
time-origin averaging.
This quantity measures the spatial fluctuations in 
$\tau_{\alpha}$. The dependence of $\chi_\tau$ on $L_B$ and $T$ (see the SI) clearly shows the presence of a length scale that grows at $T$ is decreased. Since the system-size dependence of $\tau_\alpha$ itself (for systems with periodic boundary conditions) is governed by the static length scale $\xi_s$, one can expect the system-size dependence of $\chi_\tau$ also to be controlled by the same length scale. To check whether this is true,
we performed a scaling analysis to find the length scale $\xi$ that leads to a scaling collapse of the data for $\chi_\tau(L_B,T)$. We find 
good data collapse, as shown in Fig.(\ref{tauChi}). The temperature dependence of the length scale obtained from the scaling collapse is 
found to be very similar to that of the static length scale $\xi_s$ obtained in earlier work (see the inset of Fig.(\ref{tauChi})). This 
result, which shows that the block-size dependence of $\chi_\tau(L_B,T)$ is indeed governed by $\xi_s$, is very useful as it shows that the 
static length scale can be
extracted from experimental or simulation data obtained for a single system of moderately large size. 
Similar analysis done for the 3dR10 model system are shown in SI. For this model
also the results are in good agreement with those obtained
from conventional methods.
 

To summarize, in this work, we present an efficient method which can be
used in simulations as well as in colloidal experiments on glass forming liquids
to obtain both static and dynamic length scales. Our results are validated from 
comparisons with those of conventional methods. This method has the advantage of
capturing all the important fluctuations in the system which is not possible in  simulations 
in the canonical ensemble for varying system sizes. Block analysis also provides 
extremely well-averaged results without any additional computational overhead 
in simulations and it can be implemented without much difficulty in colloidal 
glass experiments. We hope that this work will motivate experiments on colloidal glasses to measure these length
scales.
\bibliography{blockAnalysisV3}

\begin{thebibliography}{10}

\bibitem{book1}
P.~Debenedetti, {\em Metastable Liquids}.
\newblock Princeton University Press, Princeton, New Jerse, 1997.

\bibitem{book2}
P.~G. Wolynes and V.~Lubchenko, {\em Structural Glasses and Supercooled
  Liquids: Theory, Experiment, and Applications}.
\newblock John Wiley \& Sons, 2012.

\bibitem{rev1}
S.~Karmakar, C.~Dasgupta, and S.~Sastry, ``Growing length scales and their
  relation to timescales in glass-forming liquids,'' {\em Annu. Rev. Condens.
  Matter Phys.}, vol.~5, pp.~255--284, Mar 2014.

\bibitem{rev2}
S.~Karmakar, C.~Dasgupta, and S.~Sastry, ``Length scales in glass-forming
  liquids and related systems: a review,'' {\em Reports on Progress in
  Physics}, vol.~79, p.~016601, Dec 2015.

\bibitem{11BB}
L.~Berthier and G.~Biroli, ``Theoretical perspective on the glass transition
  and amorphous materials,'' {\em Rev. Mod. Phys.}, vol.~83, pp.~587--645, Jun
  2011.

\bibitem{ARPC58}
V.~Lubchenko and W.~G. Peter, ``Theory of structural glasses and supercooled
  liquids,'' {\em Annual Review of Physical Chemistry}, vol.~58, pp.~235--266,
  Oct 2006.

\bibitem{PhysRevLett.97.195701}
G.~Biroli, J.-P. Bouchaud, K.~Miyazaki, and D.~R. Reichman, ``Inhomogeneous
  mode-coupling theory and growing dynamic length in supercooled liquids,''
  {\em Phys. Rev. Lett.}, vol.~97, p.~195701, Nov 2006.

\bibitem{ktw4}
T.~R. Kirkpatrick, D.~Thirumalai, and P.~G. Wolynes, ``Scaling concepts for the
  dynamics of viscous liquids near an ideal glassy state,'' {\em Phys. Rev. A},
  vol.~40, pp.~1045--1054, Jul 1989.

\bibitem{JCP121-7347}
J.-P. Bouchaud and G.~Biroli, ``On the
  adam-gibbs-kirkpatrick-thirumalai-wolynes scenario for the viscosity increase
  in glasses,'' {\em The Journal of Chemical Physics}, vol.~121,
  pp.~7347--7354, July 2004.

\bibitem{PNASUSA2009}
S.~Karmakar, C.~Dasgupta, and S.~Sastry, ``Growing length and time scales in
  glass-forming liquids,'' {\em Proc. Natl. Acad. Sci. USA}, vol.~106,
  pp.~3675--3679, Jan 2009.

\bibitem{PhysRevLett.105.015701}
S.~Karmakar, C.~Dasgupta, and S.~Sastry, ``Analysis of dynamic heterogeneity in
  a glass former from the spatial correlations of mobility,'' {\em Phys. Rev.
  Lett.}, vol.~105, p.~015701, Jul 2010.

\bibitem{IT}
I.~Tah, S.~Sengupta, S.~Sastry, C.~Dasgupta, and S.~Karmakar, ``Glass
  transition in supercooled liquids with medium range crystalline order,'' {\em
  arXiv:1705.09532}, 2017.

\bibitem{PhysRevLett.105.217801}
E.~Flenner and G.~Szamel, ``Dynamic heterogeneity in a glass forming fluid:
  Susceptibility, structure factor, and correlation length,'' {\em Phys. Rev.
  Lett.}, vol.~105, p.~217801, Nov 2010.

\bibitem{Nature2008}
G.~Biroli, J.-P. Bouchaud, A.~Cavagna, T.~S. Grigera, and P.~Verrocchio,
  ``Thermodynamic signature of growing amorphous order in glass-forming
  liquids,'' {\em Nat Phys}, vol.~99, pp.~771--775, Oct 2008.

\bibitem{Hocky2012}
G.~M. Hocky, T.~E. Markland, and D.~R. Reichman, ``Growing point-to-set length
  scale correlates with growing relaxation times in model supercooled
  liquids,'' {\em Phys. Rev. Lett.}, vol.~108, p.~225506, Jun 2012.

\bibitem{kobVargasBerthier}
W.~Kob, S.-R. Vargas, and L.~Berthier, ``Non-monotonic temperature evolution of
  dynamic correlations in glass-forming liquids,'' {\em Nat Phys}, vol.~8,
  pp.~164--167, Feb 2012.

\bibitem{szamel3}
E.~Flenner and G.~Szamel, ``Characterizing dynamic length scales in
  glass-forming liquids,'' {\em Nat Phys}, vol.~8, pp.~696--697, Oct 2012.

\bibitem{PhysicaA}
S.~Karmakar, E.~Lerner, and I.~Procaccia, ``Direct estimate of the static
  length-scale accompanying the glass transition,'' {\em Physica A: Statistical
  Mechanics and its Applications}, vol.~391, pp.~1001--1008, Jun 2012.

\bibitem{PhysRevLett.111.165701}
G.~Biroli, S.~Karmakar, and I.~Procaccia, ``Comparison of static length scales
  characterizing the glass transition,'' {\em Phys. Rev. Lett.}, vol.~111,
  p.~165701, Oct 2013.

\bibitem{CTPRE12}
P.~Charbonneau and G.~Tarjus, ``Decorrelation of the static and dynamic length
  scales in hard-sphere glass formers,'' {\em Phys. Rev. E}, vol.~87,
  p.~042305, Apr 2013.

\bibitem{annurev.physchem.51.1.99}
M.~D. Ediger, ``Spatially heterogeneous dynamics in supercooled liquids.,''
  {\em Annual Review of Physical Chemistry}, vol.~51, pp.~99--128, Oct 2000.

\bibitem{colloid}
E.~R. Weeks, J.~C. Crocker, A.~C. Levitt, A.~Schofield, and D.~A. Weitz,
  ``Three-dimensional direct imaging of structural relaxation near the
  colloidal glass transition,'' {\em Science.}, vol.~287, pp.~627--631, Jan
  2000.

\bibitem{kegel}
W.~K. Kegel and A.~V. Blaaderen, ``Direct observation of dynamical
  heterogeneities in colloidal hard-sphere suspensions,'' {\em Science.},
  vol.~287, pp.~290--293, Jan 2000.

\bibitem{05Berthier}
L.~Berthier, G.~Biroli, J.-P.~C. Bouchaud, L.~Masri, D.~E.~L. ˇote, D.~Ladieu,
  F.~Pierno, and M.~Pierno, ``Direct experimental evidence of a growing length
  scale accompanying the glass transition,'' {\em Science}, vol.~310,
  pp.~1797--1800, Jun 2005.

\bibitem{dhlengthexp2}
C.~Dalle-Ferrier, C.~Thibierge, C.~Alba-Simionesco, L.~Berthier, G.~Biroli,
  J.-P. Bouchaud, F.~Ladieu, D.~L'H\^ote, and G.~Tarjus, ``Spatial correlations
  in the dynamics of glassforming liquids: Experimental determination of their
  temperature dependence,'' {\em Phys. Rev. E}, vol.~76, p.~041510, Oct 2007.

\bibitem{fifthorder}
S.~Albert, T.~h. Bauer, M.~Michl, B.~G, J.-P. Bouchaud, A.~Loidl,
  P.~Lunkenheime, R.~Tourbot, W.~Gasquet, C, and F.~Ladieu, ``Fifth-order
  susceptibility unveils growth of thermodynamic amorphous order in
  glass-formers,'' {\em Science}, vol.~10, pp.~1308--1311, Jun 2016.

\bibitem{PhysRevLett.116.098302}
B.~Zhang and X.~Cheng, ``Structures and dynamics of glass-forming colloidal
  liquids under spherical confinement,'' {\em Phys. Rev. Lett.}, vol.~116,
  p.~098302, Mar 2016.

\bibitem{rajesh}
K.~H. Nagamanasa, S.~Gokhale, A.~k. Sood, and R.~Ganapathy, ``Direct
  measurements of growing amorphous order and non-monotonic dynamic
  correlations in a colloidal glass-former,'' {\em Nat Phys}, vol.~11,
  pp.~403--408, May 2015.

\bibitem{cdg1}
C.~Dasgupta, A.~V. Indrani, S.~Ramaswamy, and M.~K. Phani, ``Is there a growing
  correlation length near the glass transition?,'' {\em EPL (Europhysics
  Letters)}, vol.~15, p.~303, Feb 1991.

\bibitem{fss}
K.~Binder, ``Finite size scaling analysis of ising model block distribution
  functions,'' {\em Physik B - Condensed Matter}, vol.~43, p.~119, Apr 1981.

\bibitem{blocking}
K.~E. Avila, H.~E. Castillo, A.~Fiege, K.~Vollmayr-Lee, and A.~Zippelius,
  ``Strong dynamical heterogeneity and universal scaling in driven granular
  fluids,'' {\em Phys. Rev. Lett.}, vol.~113, p.~025701, Jul 2014.

\bibitem{KA}
W.~Kob and H.~C. Andersen, ``Testing mode-coupling theory for a supercooled
  binary lennard-jones mixture i: The van hove correlation function,'' {\em
  Phys. Rev. E}, vol.~51, pp.~4626--4641, May 1995.

\bibitem{R10}
S.~Karmakar, E.~Lerner, I.~Procaccia, and J.~Zylberg, ``Statistical physics of
  elastoplastic steady states in amorphous solids: Finite temperatures and
  strain rates,'' {\em Phys. Rev. E}, vol.~82, p.~031301, Sep 2010.

\bibitem{IPL}
U.~R. Pedersen, T.~B. Schroder, and J.~C. Dyre, ``Repulsive reference potential
  reproducing the dynamics of a liquid with attractions,'' {\em Phys. Rev.
  Lett.}, vol.~105, p.~157801, Oct 2010.

\end{thebibliography}


\begin{thebibliography}{1}

\bibitem{KA}
W.~Kob and H.~C. Andersen, ``Testing mode-coupling theory for a supercooled
  binary lennard-jones mixture i: The van hove correlation function,'' {\em
  Phys. Rev. E}, vol.~51, pp.~4626--4641, May 1995.

\bibitem{R10}
S.~Karmakar, E.~Lerner, I.~Procaccia, and J.~Zylberg, ``Statistical physics of
  elastoplastic steady states in amorphous solids: Finite temperatures and
  strain rates,'' {\em Phys. Rev. E}, vol.~82, p.~031301, Sep 2010.

\bibitem{IPL}
U.~R. Pedersen, T.~B. Schroder, and J.~C. Dyre, ``Repulsive reference potential
  reproducing the dynamics of a liquid with attractions,'' {\em Phys. Rev.
  Lett.}, vol.~105, p.~157801, Oct 2010.

\bibitem{prl93.105502}
P.~Bordat, F.~Affouard, M.~Descamps, and K.~L. Ngai, ``Does the interaction
  potential determine both the fragility of a liquid and the vibrational
  properties of its glassy state?,'' {\em Phys. Rev. Lett.}, vol.~93,
  p.~105502, Sep 2004.

\bibitem{KLPZ2011}
S.~Karmakar, E.~Lerner, I.~Procaccia, and J.~Zylberg, ``Effect of the
  interparticle potential on the yield stress of amorphous solids,'' {\em Phys.
  Rev. E}, vol.~83, p.~046106, Apr 2011.

\bibitem{YDWPNAS2013}
L.~Yan, G.~During, and M.~Wyart, ``Why glass elasticity affects the
  thermodynamics and fragility of supercooled liquids,'' {\em Proc. Natl. Acad.
  Sci. USA}, vol.~110, pp.~6307--6312, Jan 2013.

\bibitem{KDSPRL16}
S.~Karmakar, C.~Dasgupta, and S.~Sastry, ``Short-time beta relaxation in
  glass-forming liquids is cooperative in nature,'' {\em Phys. Rev. Lett.},
  vol.~116, p.~085701, Feb 2016.

\bibitem{DCK16}
R.~Das, S.~Chakrabatry, and S.~Karmakar, ``A novel method to study growth of
  amorphous order in glass-forming liquids,'' {\em arXiv:1608.01474}, 2016.

\end{thebibliography}
\bibliographystyle{ieeetr}
\end{document}


\title{Block Analysis for the Calculation of Dynamic and Static Length Scales in Glass-Forming Liquids : Supplementary Information}
\author{Saurish Chakrabarty$^{1}$}\thanks{These authors contributed equally.}
\author{Indrajit Tah$^{2}$}\thanks{These authors contributed equally.}
\author{Smarajit Karmakar$^{2}$}
\author{Chandan Dasgupta$^{3,4}$}
\affiliation{
  $^1$ International Centre for Theoretical Sciences,
  Tata Institute of Fundamental Research, 
  Shivakote, Hesaraghatta, Hubli, Bangalore, 560089, India,\\
  $^2$ Centre for Interdisciplinary Sciences,
  Tata Institute of Fundamental Research, 
  21 Brundavan Colony, Narisingi, Hyderabad, 500075, India,
  $^3$ Centre for Condensed Matter Theory, Department of Physics,
  Indian Institute of Science, Bangalore, 560012, India, 
  $^4$ Jawaharlal Nehru Centre for Advanced 
  Scientific Research, Bangalore 560064, India.}

\maketitle

\section{Models and Simulation Details}
\label{modelsAndSim}
We have studied three different model glass forming liquids in three 
dimensions. The model details are given below:
\vskip +0.3cm
\noindent{\textbf{3dKA Model:}} 

The KABLJ \cite{KA} liquid is an 80:20 mixture of two LJ 
particles, A and B. This model was first introduced by Kob-Anderson to 
simulate $Ni_{80}P{20}$. This liquid has become the well known model system 
for studying viscous liquid dynamics, like other binary mixtures
it's size ratio lies below 0.9 so it does not  crystallize easily. 

\noindent The interaction potential is given by
\[ V_{\alpha\beta}(r) = 4.0\epsilon_{\alpha\beta}[(\frac{\sigma_{\alpha\beta}}{r})^{12} - (\frac{\sigma_{\alpha\beta}}{r})^6]\]
where $\alpha,\beta \in \{ A,B \}$ and $\epsilon_{AA} = 1.0$, $\epsilon_{AB} = 1.5$, $\epsilon_{BB} = 0.5$; 
$\sigma_{AA} = 1.0$, $\sigma_{AB} = 0.80$,   $\sigma_{BB} = 0.88$. The interaction potential was cut off at 
$2.50\sigma_{\alpha\beta}$ and the number density is $\rho = 1.20$. we use a quadratic polynomial to make the potential and its first two derivatives smooth at the cutoff distance. Length, energy and time scale are measured in 
units of $\sigma_{AA}, \epsilon_{AA}$ and $\sqrt{\frac{\sigma_{AA}^2}{\epsilon_{AA}}}$. For Argon these units 
corresponds to a length of $3.4 \AA$, an energy of $120Kk_{\beta}$ and time of  $3\times 10^{-13} s$. 

\vskip +0.5cm
\noindent{\textbf{3dR10 Model:}} 

This is a \textbf{50:50} binary mixture \cite{R10} 
with the pure repulsive pair wise interaction potential, defined as 
\[ V_{\alpha\beta}(r) = \epsilon_{\alpha\beta}[(\frac{\sigma_{\alpha\beta}}{r})^{10}]\]
Here $\epsilon_{\alpha\beta} = 1.0$, $\sigma_{AA} = 1.0$, $\sigma_{AB} = 1.22$,   
$\sigma_{BB} = 1.40$. The interaction potential is cut-off at 
$1.38\sigma_{\alpha\beta}$. The number density $\rho = 0.81$.

\vskip +0.5cm
\noindent{\textbf{3dIPL Model:}} 

This model glass forming liquid is a variant of 3dKA 
model with purely repulsive pairwise interaction. The model was first studied 
in \cite{IPL}. Here the interaction potential is given by 
\[ V_{\alpha\beta}(r) = 1.945\epsilon_{\alpha\beta}[(\frac{\sigma_{\alpha\beta}}{r})^{15.48}]\]
All the parameters and interaction cut-off are same as 3dKA model. Though this 
is another model with pure repulsive interaction, it is important to note that 
for this model the interaction range is much larger than the 3dR10 model. In 
3dR10 model the interaction range extends only up to first neighbour, whereas
for 3dIPL it extends up to second neighbour. The range of interaction is known
to play an important role in determining both dynamic and mechanical properties 
of the system \cite{prl93.105502,KLPZ2011,YDWPNAS2013}.  

We have done NVT simulations for all the model systems using velocity-verlet 
integration scheme. The system sizes studied are $N = 10000$ particles. 
For 3dKA model we have reported results for $N = 64000$ particles also. We showed
that the results do not change quantitatively even if we use larger system size. 
Note that the data with larger system sizes are always preferred for this 
analysis. For all the simulations we have first equilibrated our systems at 
least for $100\tau_\alpha$ and stored data for similar simulation time. We have
also performed $32$ statistically independent simulations for each temperature
and system size studied for better averaging. For $N = 64000$, we have done $16$
statistically independent simulations for ensemble averaging. 

\begin{figure*}[!htpb]
  \begin{center}
		\includegraphics[scale=0.450]{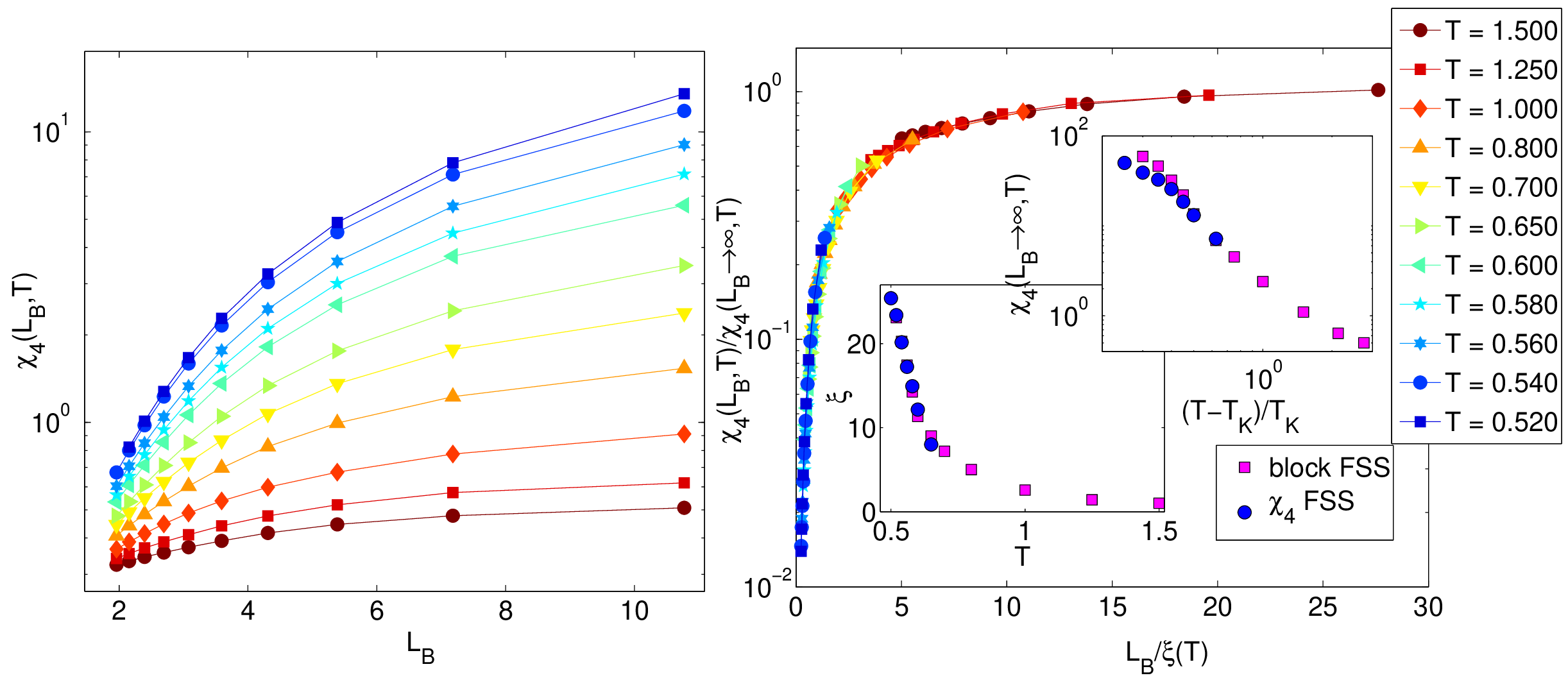}
   \includegraphics[scale=0.450]{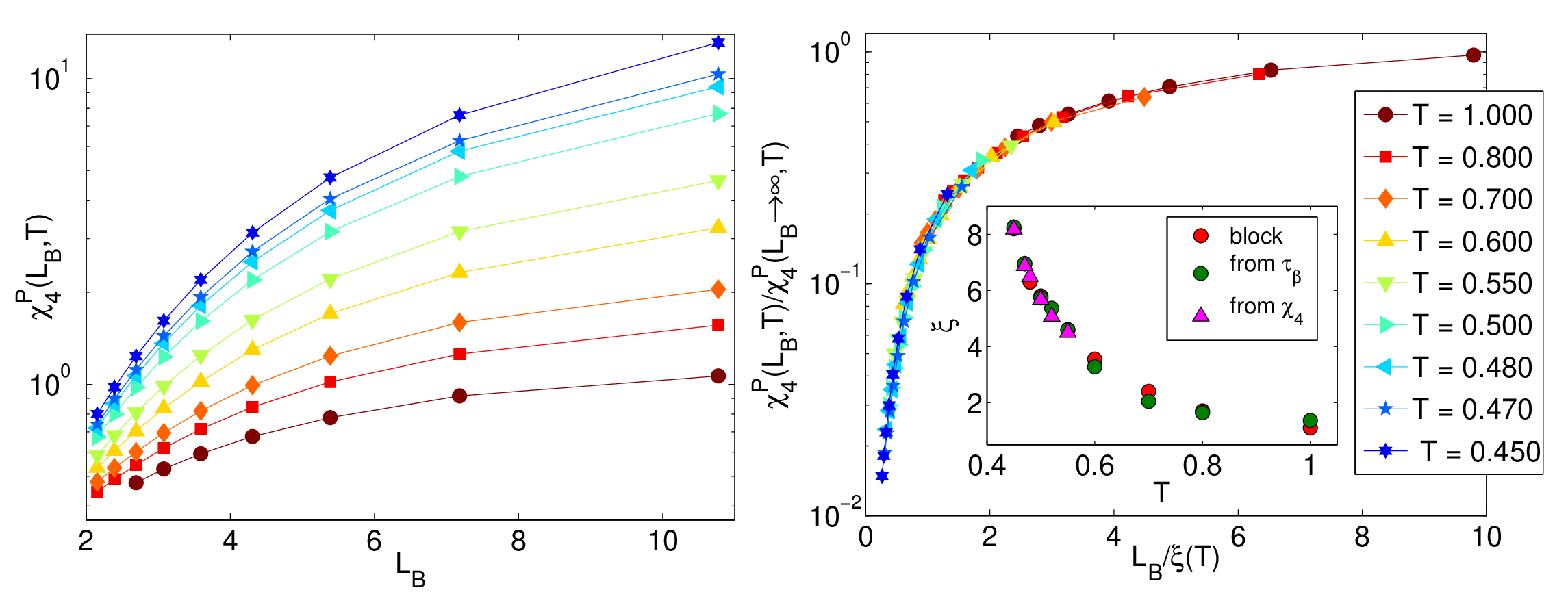}
    \caption{Top panels: Block size dependence of $\chi_4^P$ 
      for the 3dR10 model. In the left panel, we plot $\chi_4^P$ versus 
      the block size and in the right panel a collapse is done by rescaling 
      the $x$-axis using a suitable length-scale, $\xi(T)$ and the $y$-axis 
      using the saturation value of $\chi_4^P$ for infinite block size.
      In the inset, the temperature dependences of $\xi(T)$ and
      $\chi_4^P(\infty,T)$ are compared with the corresponding quantities 
      obtained using conventional finite-size-scaling. We have used 
      $T_K \simeq 0.40$ for this model. Bottom Panels: Similar analysis 
      done for 3dIPL model. Here in addition, the length scale obtained 
      from the collapse of the $\beta$-relaxation time \cite{KDSPRL16} is 
      also compared with the above length scales.}
    \label{chi4BlockAnalysis3dIPL}
  \end{center}
\end{figure*}

\section{Block size dependence of $\chi_4^P$ for the 3dR10 and 3dIPL models.}
To confirm the results reported for the block size dependence of $\chi_4^P$
for 3dKA model, we have done similar analysis for 3dR10 and 3dIPL models.
The results are shown in Fig.\ref{chi4BlockAnalysis3dIPL}. Top panels of 
Fig.\ref{chi4BlockAnalysis3dIPL} show the system size dependence of $\chi_4^P$ 
for the 3dR10 model and the scaling collapse. Bottom panels of the same 
figure show the results obtained for the 3dIPL model. The scaling observed 
for both these model systems are very good and the obtained
length scales from the block FSS analysis are in very good agreement with 
the results obtained by other conventional methods. For the 3dIPL model 
(shown in lower panels of the same figure) the length scale obtained from 
the FSS of short time scale, $\tau_{\beta}$ (taken from \cite{KDSPRL16}) 
are also compared with the length scale obtained from the block analysis. 
In \cite{KDSPRL16}, it is shown that the FSS of $\tau_{\beta}$ are also 
controlled by the dynamic heterogeneity length scale.

\section{Block size dependence of $\chi_\tau$}
\begin{figure*}[!htpb]
\begin{center}
\includegraphics[scale=0.60,angle=90]{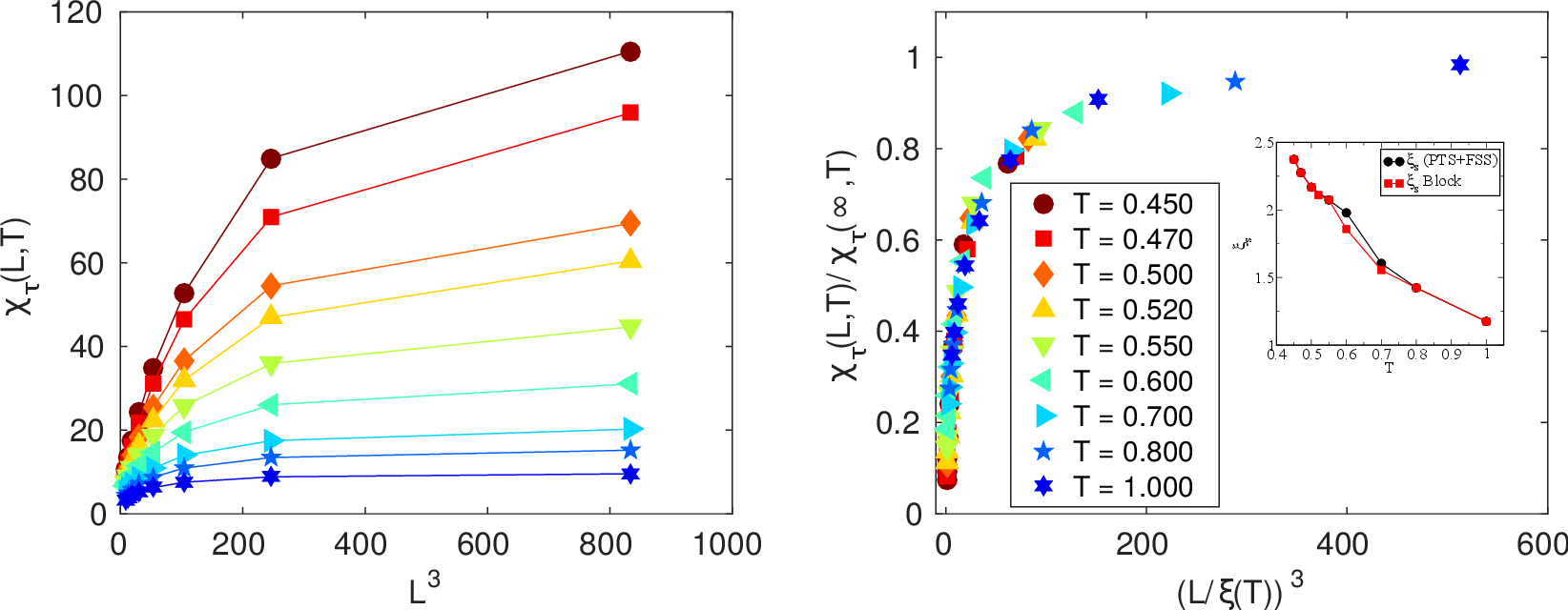}
\vskip +0.5cm
\includegraphics[scale=0.60,angle=90]{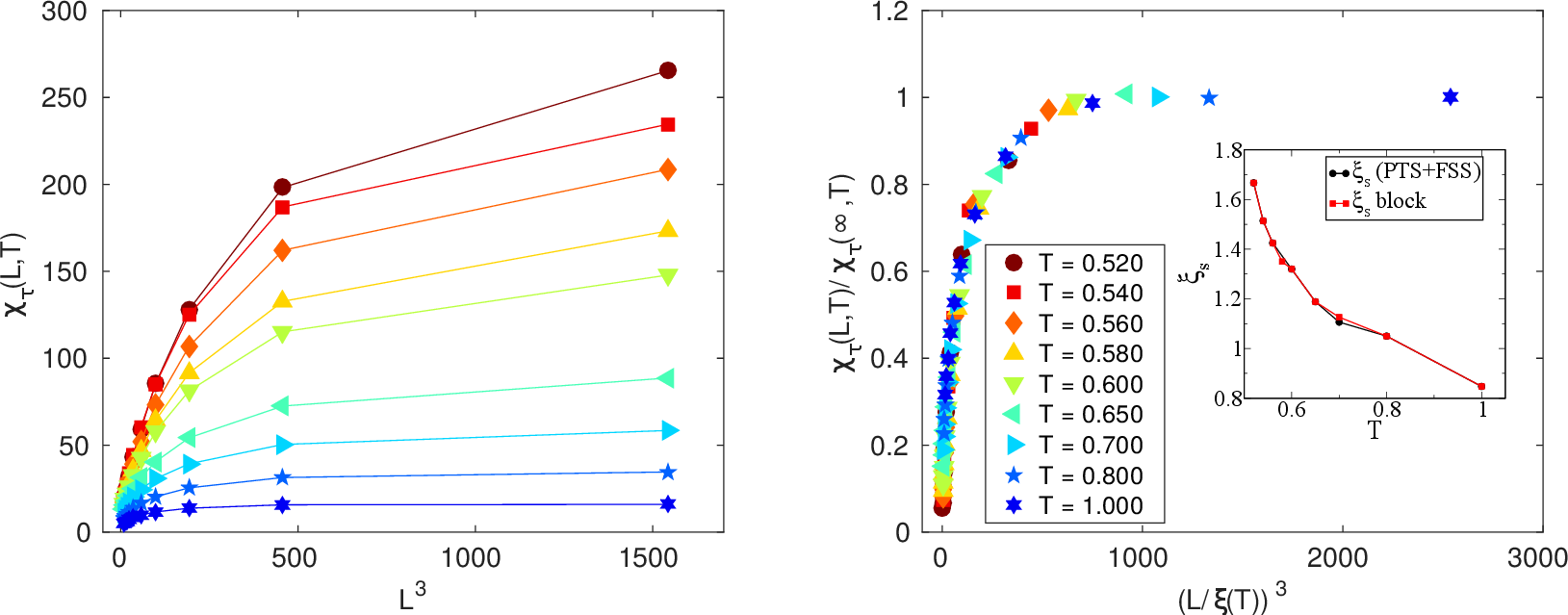}
\caption{Top panels: Block size dependence of $\chi_\tau$ for the 3dKA model. 
In the left panel, we plot $\chi_\tau$ versus the block size and in the right 
panel a collapse is done by rescaling the $x$-axis using a suitable length-scale,
$\xi(T)$ and the $y$-axis using the saturation value of $\chi_\tau$ for 
infinite block size. In the inset, the temperature dependence of $\xi(T)$ is
compared with the static length scale obtained using conventional finite-size-
scaling and PTS method. Bottom panels: Similar analysis done for 3dR10 model.}
\label{chiTauBlockAnalysis}
\end{center}
\end{figure*}

In the main article, we have discussed about the scaling collapse of $\chi_\tau$
using the static length scale for 3dKA model. Here in 
Fig.\ref{chiTauBlockAnalysis}, we have shown the $\chi_\tau$
for different block sizes for different temperatures for both 3dKA and 3dR10 model. 
Similar to 3dKA model, FSS of $\chi_\tau$ for 3dR10 model is also controlled by the 
corresponding static length scale of 3dR10 model. This establishes the usefulness 
of block analysis for extracting both static and dynamic length scales in a system
with relatively less computational or experimental efforts. ``FSS+PTS'' refers
to the static length scale obtained using FSS of $\alpha$-relaxation time and 
Point-to-Set (PTS) method in cavity geometry. The data is taken from 
Ref.\cite{DCK16}.

\bibliography{block_supply}
\bibliographystyle{ieeetr}